\documentclass[aps,prl,twocolumn,floatfix,showpacs,preprintnumbers,groupedaddress]{revtex4}

\usepackage{graphics}
\usepackage{graphicx}
\usepackage{epsfig}

\newcommand{\beq}{\begin{equation}}
\newcommand{\eeq}{\end{equation}}
\newcommand{\bea}{\begin{eqnarray}}
\newcommand{\eea}{\end{eqnarray}}
\usepackage{dcolumn}% Align table columns on decimal point
\usepackage{bm}% bold math%
\begin{document}
\title{Dynamics of Pinned Magnetic Vortices}
\author{R. L. Compton}
\affiliation{School of Physics and Astronomy, University of Minnesota, 116 Church St. SE\\
Minneapolis, MN 55455}
\author{P. A. Crowell}
  \email[]{crowell@physics.umn.edu}
\affiliation{School of Physics and Astronomy, University of Minnesota, 116 Church St. SE\\
Minneapolis, MN 55455}

\begin{abstract}
We observe the dynamics of a single magnetic vortex in the presence of a random array of pinning sites.  At low excitation amplitudes, the vortex core gyrates about its equilibrium position with a frequency that is characteristic of a single pinning site. At high amplitudes, the frequency of gyration is determined by the magnetostatic energy of the entire vortex, which is confined in a micron-scale disk. We observe a sharp transition between these two amplitude regimes that is due to depinning of the vortex core from a local defect.  The distribution of pinning sites is determined by mapping fluctuations in the frequency as the vortex core is displaced by a static in-plane magnetic field.

\end{abstract}


\maketitle The excitation spectrum of a sub-micron ferromagnetic particle is influenced profoundly by its shape.  An example of considerable interest is the vortex state of a soft ferromagnetic disk, in which the lowest frequency excitation is a translational mode of the vortex core.  This mode, in which the vortex core gyrates about its equilibrium position, has been studied in several recent experiments \cite{ParkCrowell:PRB2003,Choe:Science2004,Back:PRL2005,Novosad:PRB2005,Guslienko:PRL2006,ZhuFreeman:PRB2005}.  Although the core has a diameter of the order of the exchange length ($\sim$10~nm) \cite{Guslienko:PRB2001,Wachowiak:Science2002}, the eigenfrequency of the gyrotropic mode is determined only by the geometrical aspect ratio (diameter over thickness) of the disk in which it is confined \cite{Guslienko:JAP2002,IvanovZaspel:JAP2004}.  In particular, the gyrotropic frequency should be independent of the location of the vortex core in the disk \cite{Novosad:PRB2005}.  As demonstrated recently, however, the vortex core can be pinned by defects in a patterned thin film \cite{Shima:JAP2002,RahmWeiss:JAP2004,UhligZweck:PRL2005}.  In the case of intrinsic defects, the range of the effective pinning potential appears to be on the order of ten nanometers \cite {UhligZweck:PRL2005}.  This raises the question of how a nanoscale defect influences the vortex gyrotropic motion.

In this Letter, we report that pinning has a pronounced effect on magnetic vortex dynamics at small amplitudes.  The vortex gyrotropic frequency in 50~nm thick permalloy (Py) disks with diameters from 600~nm to 2~$\mu$m fluctuates by a factor of at least two as the core is displaced by a static magnetic field over length scales $\sim$~10 nm.  The dependence on core location indicates that the small-amplitude dynamics are influenced strongly by the characteristics of a particular pinning site.  Each site has a critical excitation amplitude above which the vortex becomes depinned and gyrates at the frequency determined by the aspect ratio of the disk.  We image the spatial distribution of pinning sites in a disk by displacing the vortex core in two dimensions while monitoring the gyrotropic frequency.  Using this approach, we find a density of up to $\sim~2\times10^{11}$~cm$^{-2}$ for the pinning sites in our films.  Although this density cannot be correlated directly with the physical morphology of the permalloy films, the extreme sensitivity to core displacement indicates that the magnetic microstructure fluctuates strongly on nanometer scales.

The samples were patterned from 50 nm thick films of Py (Ni$_{0.81}$Fe$_{0.19}$), which has negligible magnetocrystalline anisotropy.  To enhance pinning, crystallographically textured films were sputtered at a base pressure of $5\times10^{-8}$ Torr onto a buffer/ substrate system of 30 nm Cu(111)/ Al$_{2}$O$_{3}$(0001), in which the grain size depends on substrate temperature \cite{LundLeighton:JVacSci2004}.  The films were capped at room temperature with 2.5 nm Al and 50 nm (optically transparent) SiN.  Results will be presented for two samples.  A small-grain (SG) sample was grown at $30^{\circ}$C, yielding an average grain diameter of $\approx$~35 nm as determined by atomic force microscopy.  A large-grain (LG) sample with grain diameter $\approx$~85 nm was grown at $200^{\circ}$C. The average roughness (R$_{A}$)is $\approx$~4 \AA~(SG) and $\approx$~8 \AA~(LG). Wide angle X-ray diffraction indicates (111) texture and low strain. Patterning was performed by electron beam lithography and dry etching.

\begin{figure}
    \centerline{\epsfbox{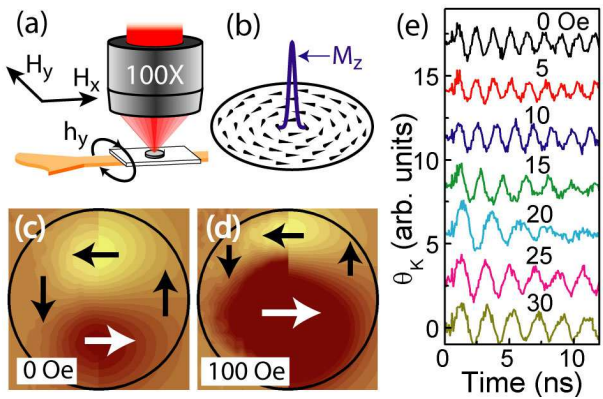}}
    \caption{(color online) (a) Schematic of the experiment. (b) Schematic of a vortex. The $z$-component (out-of-plane) of the magnetization is non-zero only in the core. (c) In-plane magnetization configuration for a 2 $\mu$m disk (arrows) and image of the $z$-component of the magnetization (shading) 80~psec after the arrival of the field pulse.  The split frame shows both experimental (left) and simulated (right)images in a field of 0 Oe.   (d) The same image in a static field of 100~Oe.  (e)  Time scans of the polar Kerr signal obtained in several different magnetic fields with the objective positioned 300~nm from the center of a 1~$\mu$m diameter disk.}
    \label{fig:fig1}
\end{figure}

We use time-resolved Kerr microscopy (TRKM) \cite{HiebertFreeman:PRL1997} to study vortex core dynamics in individual magnetic disks.  The substrate is polished to a thickness less than 30~$\mu$m, and the sample is then positioned on a coplanar waveguide as shown in Fig.~\ref{fig:fig1}(a).  The sample is excited by a magnetic field pulse with a temporal width less than 120 ps and amplitude 2 to 13~Oe (estimated from the magnitude of the current) oriented in the plane of the disk.  A linearly polarized laser pulse (duration 150 fs, $\lambda$ = 810 nm) is focused through an oil-immersion objective to a spot with a full width at half maximum (FWHM) $\sim$~400 nm.  Using the polar Kerr effect, we measure the pump-induced change in the $z$-component of the magnetization as a function of the time delay between the pump and the probe. In the vortex state, shown schematically in Fig.~\ref{fig:fig1}(b), the torque due to the pump pulse rotates the magnetization in the upper and lower halves of the disk into and out of the plane.  Fig.~\ref{fig:fig1}(c) shows a polar Kerr image of  a 2 $\mu$m diameter SG disk at the peak of the pump pulse, demonstrating the expected contrast.  The left half of this image is data and the right half is a micromagnetic simulation of the response convolved with a Gaussian of FWHM 400 nm \cite{LLG}.  An image taken in a field of 100~Oe applied in the $x$-direction is shown in Fig.~\ref{fig:fig1}(d).  As expected, the vortex core is displaced in a direction perpendicular to the field \cite{Guslienko:PRB2001,UhligZweck:PRL2005}.  Measurements of the  position $y$  of the nodal line in the image allow for a calibration of the core displacement, which is approximately linear with field.  Experimental calibrations are $dy/dH = 4.5 \pm 0.3$~nm/Oe for the 2~$\mu$m disk and $0.9 \pm 0.2$~nm/Oe for the 1~$\mu$m disk.  In comparison, micromagnetic simulations give $dy/dH = 3.9$~nm/Oe  and 1.1~nm/Oe respectively.  $dy/dH$ increases approximately quadratically with disk diameter in both experiment and simulations, although we cannot measure the displacement reliably for the smallest disks.

Time scans obtained at a radius of 300 nm from the center of a 1~$\mu$m diameter SG disk are shown in Fig.~\ref{fig:fig1}(e).  The oscillations are due to the gyrotropic motion of the vortex core about its equilibrium position \cite{Guslienko:JAP2002,ParkCrowell:PRB2003,Choe:Science2004,Novosad:PRB2005,Guslienko:PRL2006}.  The sub-GHz gyrotropic frequency $\omega_0$ and damping times greater than 10~ns observed in Fig.~\ref{fig:fig1}(e) are typical \cite{ParkCrowell:PRB2003,Choe:Science2004,ZhuFreeman:PRB2005}, but the dependence of the frequency on applied field is unexpected.  In the case of an ideal vortex \cite{Guslienko:JAP2002}, $\omega_0 = \gamma M_s \xi^2/2\chi$, where $\gamma$ is the gyromagnetic ratio, $M_s$ is the saturation magnetization, $\xi$ is a constant reflecting the boundary conditions, and $\chi = dM/dH$ is the susceptibility, which is nearly independent of field \cite{Shima:JAP2002}.  The gyrotropic frequency should therefore show very weak dependence on magnetic field, as confirmed in a recent resonance experiment \cite{Novosad:PRB2005}.

\begin{figure}
    \centerline{\epsfbox{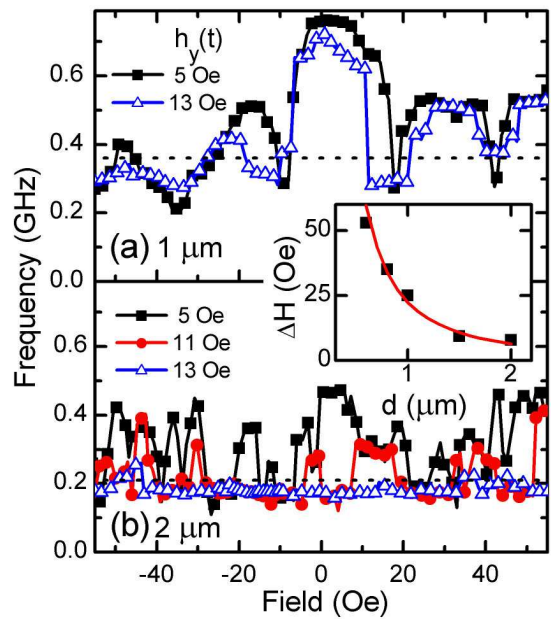}}
    \caption{(color online) Field dependence of the vortex gyrotropic frequency for $1 \mu$m (a) and $2 \mu$m (b) diameter disks fabricated from SG films.  The amplitudes of the excitation pulse are 5 Oe (black squares), 11 Oe (red circles), and 13 Oe (blue triangles).  The dotted lines are the expected frequencies from micromagnetic simulations. Inset:  The average period of the fluctuations (black squares) vs disk diameter.  The solid line is a model (see text) based on a defect spacing of 25 nm.}
    \label{fig:fig2}
\end{figure}

We have also investigated the field dependence of $\omega_0$ for different excitation amplitudes.  The Fourier transform of the time-resolved data at each field are fit to a Lorentzian to obtain $\omega_0$.  Fig. \ref{fig:fig2} shows the dependence of $\omega_0$ on static field for 1$\mu$m and 2$\mu$m diameter SG disks. Each set of data corresponds to a different pump pulse amplitude $h_y$.  For small pump pulse amplitudes, the observed gyrotropic frequency fluctuates irregularly, with frequency peaks that are twice the expected field-independent value (dotted lines in Fig.~\ref{fig:fig2}) \cite{Guslienko:JAP2002,IvanovZaspel:JAP2004}.  The fluctuations are highly repeatable for a given disk, but are uncorrelated between neighboring disks.  The data are independent of the vortex polarity (orientation of the vortex core into or out of the plane) but the field dependence is inverted upon reversal of the chirality (curling direction of the in-plane magnetization).  The frequency fluctuates with a characteristic period $\Delta H$ = 23 Oe (obtained by counting frequency peaks over a field range of hundreds of Oe) for the 1 $\mu$m disk and 7 Oe for the 2 $\mu$m disk.  $\Delta H$ measured for different disk diameters is shown as symbols in the inset of Fig.~\ref{fig:fig2}.  Note that $\Delta H$ increases by approximately a factor of 7 as the disk diameter decreases from 2~$\mu$m to 600~nm.

\begin{figure}
    \centerline{\epsfbox{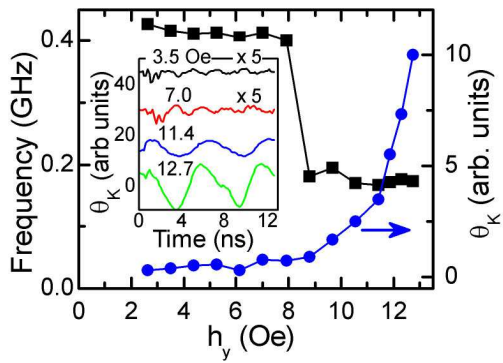}}
    \caption{(color online) The vortex gyrotropic frequency (black squares, left axis) for the 2~$\mu$m disk is shown as a function of field pulse amplitude at a fixed static field of -23 Oe.  The maximum magnitude of the gyrotropic response (blue circles, right axis) increases rapidly for pulse amplitudes above 8~Oe.  Time-resolved data at several different excitation pulse amplitudes are shown in the inset.}
    \label{fig:fig3}
\end{figure}

As shown in Fig.~\ref{fig:fig2}, an excitation pulse amplitude of 13~Oe slightly suppresses the field-dependent frequency fluctuations for the 1 $\mu$m disk and almost completely eliminates the fluctuations for the 2 $\mu$m disk. The dependence of the magnitude of the response on pulse amplitude is shown for the 2~$\mu$m disk in Fig.~\ref{fig:fig3}.  For this measurement the static field is held fixed at one of the frequency peaks of Fig. \ref{fig:fig2}(b) while the excitation pulse amplitude is varied.  The most important feature of Fig.~\ref{fig:fig3} is the rapid drop in the gyrotropic frequency from 0.4 to 0.2 GHz at a pulse amplitude of 8~Oe. The magnitude of the response increases rapidly above this critical pulse amplitude.

The important features of Figs.~\ref{fig:fig2} and~\ref{fig:fig3} are consistent with the existence of point-like defects in the Py disk that pin the vortex core.  As shown in a recent Lorentz microscopy experiment  \cite{UhligZweck:PRL2005}, the core moves from defect to defect as the magnetic field is varied.  Fig.~\ref{fig:fig2} indicates that movement of the core among pinning sites leads to fluctuations in the gyrotropic frequency.  Different disks fabricated from the same film should have an identical areal defect density $n$, and therefore the fluctuations should be associated with a single length scale.   To test this hypothesis, we use the calibrations $dy/dH$ for the core displacement determined from micromagnetic simulations for each disk diameter.  As noted above, these agree well with the experimental calibrations measured for the larger disks.  For each diameter $d$, we determine the field scale $\Delta H = \Delta y (dy/dH)^{-1}$ required to translate the vortex core over a distance $\Delta y = \sqrt{(1/n)}$. The results for $\Delta y= 25$~nm are shown as the solid curve in the inset of Fig.~\ref{fig:fig2}.  The good agreement with experiment demonstrates that the different periods $\Delta H$ for the different diameters are associated with a single length scale characteristic of the defect density in the Py film.

The second important feature of the experimental data is the existence of a critical pulse amplitude above which the gyrotropic frequency decreases abruptly to the value expected given the magnetostatic potential of the disk \cite{Guslienko:JAP2002}.  The critical amplitude corresponds to the depinning of the vortex core from a defect, which occurs when the radius of the gyrotropic orbit exceeds the range of the pinning center.  At the maximum pulse amplitude, we are able to achieve depinning for almost all sites in disks with diameters greater than 1.5~$\mu$m.  Above the depinning transition, the dynamics should be governed by the Thiele force equation \cite{Thiele:PRL1973,Huber:PRB1982,Guslienko:JAP2002}
\begin{equation}
    \mathbf{G} \times \frac{d\mathbf{X}}{dt} - k \mathbf{X} -D \frac{d\mathbf{X}}{dt} = \mathbf{F}(t),
\label{eq:thiele}
\end{equation}
where $\mathbf{G} = -G\hat{z}$ is the vortex gyrovector, $\mathbf{X}$ is the displacement of the vortex core from its equilibrium position, $k$ is a stiffness determined from the magnetostatic energy of the displaced vortex, and $D$ is a damping constant. $\mathbf{F}(t)\perp \mathbf{h}(t)$ is the time dependent excitation force, with $\mathbf{h}(t)$ the excitation field. The resonant gyrotropic frequency obtained from the solution of Eq.~\ref{eq:thiele} is $\omega_0 =k/G$ \cite{frequency}, and the maximum response amplitude is proportional to $1/\omega_0 D$.  It has been shown elsewhere that the solutions of the Thiele equation provide an excellent description of the gyrotropic mode in permalloy disks \cite{ParkCrowell:PRB2003,Novosad:PRB2005}.  For the 2 $\mu$m disk, we find $\omega_{0}$ is independent of static magnetic field for pulse amplitudes $>$10 Oe, which corresponds to a gyrotropic orbital radius $>$~10 nm.  The observed field independence in this high amplitude regime is consistent with the results of Novosad {\it et al.}\cite{Novosad:PRB2005}.

Except for the higher frequency, the vortex gyrotropic motion in the pinned case appears to be qualitatively similar to the unpinned regime.  We therefore consider the possibility that Eq.~\ref{eq:thiele} also applies for pinned vortices, but with a higher effective spring constant $k$ \cite{Pereira:PRB2005}.  For the case shown in Fig.~\ref{fig:fig3}, the change in $\omega_0$ at the critical pulse amplitude implies that $k$ in the pinned case is larger by a factor of 2.   Changes in $k$ alone, however, are not sufficient to account for the rapid increase in the magnitude of the response above the critical pulse amplitude.  Furthermore, we do not observe a clear trend in the damping times for pinned and unpinned vortices, and it is unlikely that $G$, which is independent of the size and shape of the vortex core, can change sufficiently to account for the observed increase in the gyrotropic signal.  We therefore do not have a complete explanation of the dramatic increase in the signal above the critical pulse amplitude.  Interestingly, the signals at high amplitudes are noticeably non-sinusoidal.  This is probably due to deformations in the orbit caused by nearby pinning sites and not by any intrinsic non-linearity.

\begin{figure}
    \centerline{\epsfbox{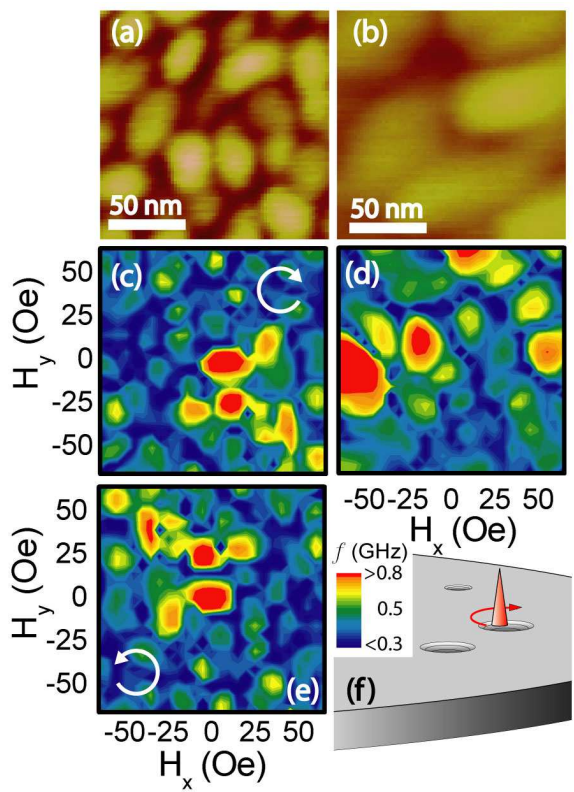}}
    \caption{(color online) (a),(b) Representative atomic force microscopy images of the SG (a) and LG (b) samples.  (c),(d) Contour maps of the gyrotropic frequency as a function of static field applied along both in-plane directions for 1~$\mu$m diameter SG (c) and LG (d) samples respectively, obtained at a pulse amplitude of $\approx$~7 Oe.  (e)  Contour map of the gyrotropic frequency for the SG sample with the vortex chirality reversed.  The image is almost identical to (c) after rotation by  180$^{\circ}$.  (f) Scale bar for frequency in (c) - (e) along with a representation of the vortex core in a random array of defects.}
    \label{fig:fig4}
\end{figure}

By applying orthogonal in-plane magnetic fields, it is possible to move the vortex core anywhere in the plane of the disk.  The core thereby functions as a nanoscale scanning probe, and the local value of the gyrotropic frequency can be used to map the pinning potential.  We have used this approach to scan 1~$\mu$m diameter SG and LG samples using a pulse amplitude of $\approx$~7 Oe.  Typical atomic force micrographs (AFM's) of the SG and LG films are shown in Figs.~\ref{fig:fig4}(a) and (b).  Magnetic fields in the $x$ and $y$ directions are varied in steps of 5 Oe over a range of 125~Oe, which leads to a core displacement over a spatial region 140~nm $\times$ 140~nm (the size of the AFM images). Maps of the gyrotropic frequency for 1~$\mu$m disks of each film are shown in Figs.~\ref{fig:fig4}(c) and (d).  The pinning sites appear as regions of high frequency. Changing the vortex chirality, as for the case of Fig.~\ref{fig:fig4}(e), reverses the direction of core displacement for a given field.  The image of Fig.~\ref{fig:fig4}(e) is therefore identical to that of Fig.~\ref{fig:fig4}(b) after inversion about the origin.

From the images of Fig.~\ref{fig:fig4}, we infer the density of pinning defects to be $\approx 1.7\times10^{11}$~cm$^{-2}$ for the SG sample and $\approx 1.0\times10^{11}$~cm$^{-2}$ for the LG sample.  The areal density of grains in Figs. \ref{fig:fig4}(a) and (b) is $\approx 8\times10^{10}$~cm$^{-2}$ (SG) and $\approx 1.4\times10^{10}$~cm$^{-2}$ (LG).  Although the density of pinning sites decreases with increasing grain size, there does not appear to be a direct correlation between the grain structure and the location of pinning sites.  One might expect the strongest pinning sites to be at grain boundaries, but the regions of high frequency in Figs.~\ref{fig:fig4}(c) and (d) appear point-like rather than filamentary.  Using the core displacement calibration $dy/dH$, the typical range of a pinning potential in the SG sample is approximately 10~nm, which is of the order of the core diameter.  The point-like contrast in Fig.~4 as well as the threshold behavior of Fig.~\ref{fig:fig3} suggest that the pinning potential acts on the core itself, lowering its exchange and magnetostatic energy.  We believe that the energy associated with the in-plane spins in the vicinity of the core is nearly independent of the core position, although small fluctuations of the in-plane anisotropy energy from grain to grain have also been proposed as a source of vortex pinning \cite{UhligZweck:PRL2005}.

In summary, we have demonstrated that pinning associated with intrinsic defects has a profound influence on the dynamics of magnetic vortices.  For small amplitudes, the field-dependence of the gyrotropic mode frequency is governed by the density and strength of pinning sites.  In this manner the vortex core serves as a sensitive probe of magnetic microstructure.

This work was supported by NSF DMR 04-06029, the NSF NNIN program through the Minnesota Nanofabrication Center, and the Minnesota Supercomputer Institute. We thank Chris Leighton and Jeff Parker for assistance with sample preparation.

\end{document}